\documentclass[aps,prl,twocolumn,showkeys,groupedaddress]{revtex4}
\usepackage{graphicx}% Include figure files
\usepackage{dcolumn}% Align table columns on decimal point
\usepackage{bm}% bold math
\usepackage{hyperref}% add hypertext capabilities
\usepackage{ulem}
\begin{document}

%\preprint{APS/123-QED}

\title{Ultra-narrow and widely tunable Mn$^{2+}$ Emission from Single Nanocrystals of ZnS-CdS alloy}
\author{Abhijit Hazarika$^{1}$, Arunasish Layek$^{2}$, Suman De$^{2}$, Angshuman Nag$^{1}$, Saikat Debnath$^{3}$, Priya Mahadevan$^{3}$, Arindam Chowdhury$^{2}$, D. D. Sarma$^{*, 1, 4, 5}$}
\affiliation{$^1$ Solid State and Structural Chemistry Unit, Indian Institute of Science, Bangalore 560012, India\\$^{2}$Department of Chemistry, Indian Institute of Technology Bombay, Powai, Mumbai 400076, India, India\\$^{3}$Department of Condensed Matter Physics and Material Science,S. N. Bose National Centre for Basic Sciences, Kolkata 700098, India\\$^{4}$Council of Scientific and Industrial Research - Network of Institutes for Solar Energy (CSIR-NISE), New Delhi, India\\$^{5}$Department of Physics and Astronomy, Uppsala University, Box-516, SE-75120, Uppsala, Sweden}

\date{\today}% It is always \today, today,

\begin{abstract}

\end{abstract}

%Use showkeys class option if keyword
\maketitle

{\bf  Extensively studied Mn-doped semiconductor nanocrystals have invariably exhibited photoluminescence (PL) over a narrow energy window of width $\leq$ 149 meV in the orange-red region and a surprisingly large spectral width ($\geq$ 180 meV), contrary to its presumed atomic-like origin. Carrying out emission measurements on individual single nanocrystals and supported by ab initio calculations, we show that Mn PL emission, in fact, can (i) vary over a much wider range ($\sim$ 370 meV) covering the deep green-deep red region and (ii) exhibit widths substantially lower ($\sim$ 60-75 meV) than reported so far, opening newer application possibilities and requiring a fundamental shift in our perception of the emission from Mn-doped semiconductor nanocrystals.}

Doping of semiconductor nanocrystal hosts with a localized, essentially atomic-like and often magnetic, impurity constitutes one of the most active research fields, yielding a wide range of interesting properties ~\cite {1, 2, 3, 4, 5, 6, 7, 8, 9, 10, 11, 12}. One of the most intensely pursued properties from such systems is the extraordinarily bright photoluminescence (PL) of a variety of Mn-doped semiconductor nanocrystals (NCs) ~\cite {13, 14, 15, 16, 17, 18, 19, 20, 21, 22, 23}. There are two distinct classes of semiconductor NCs with contrasting PL properties, namely doped and undoped ones. The undoped ones possess size dependent tunable excitonic emission, which, however, has drawbacks due to self-absorption of emission by other NCs in the ensemble ~\cite {24} and its long term stability ~\cite {25}. Both these problems are avoided in the alternate route of dopant emission, where the energy following the electron-hole excitation in the host is transferred to a dopant ion and the deexcitation involves only dopant states ~\cite {13, 14}. However, the involvement of the atomic-like Mn $\it {d}$-states implies that the emission wavelength of Mn is relatively unaffected by the size of the semiconductor host, and therefore, missing the important functionality of tunability available from excitonic emission. Experimentally, it has indeed been seen that Mn emission has very limited tunability, typically $<$ 150 meV, around 585 nm independent of the size and, to a large extent even the chemical nature of the host NC ~\cite {18, 26, 27, 28}. Another intriguing aspect is that Mn $\it {d}$-emission has been invariably found to have a large spectral width ($\sim$ 200-250 meV), incompatible with an atomic like ${\it {^4}}$${\it {T}}$${\it {_1}}$ - ${\it {^6}}$${\it {A}}$${\it {_1}}$ Mn-emission. This large width has been explained in terms of coupling of Mn $\it {d}$ levels to the vibrational structure of the host ~\cite {29}, though this spectral width is considerably larger than even that ($\sim$ 120-140 meV) found in Mn-doped bulk semiconductors, like ZnS ~\cite {1, 5}. It should be noted that these two well-accepted beliefs, namely the lack of substantial tunability of Mn emission wavelength and inevitable presence of a large spectral width due to a fundamental quantum mechanical process, places serious limits on the general usefulness of Mn dopant emissions from nanostructured host materials. The importance of the present work based on spatially resolved experiments, is the proof for both these long-standing beliefs as myths, providing direct experimental evidence of emissions of wide ranging colors from Mn emission in a semiconductor NC host and of spectral line-widths that are ultra-narrow (60-75 meV), lower by a factor of about three from the narrowest width obtained in ensemble measurements. We discuss the origin and implications of these findings with the help of extensive ab initio calculations.
\begin{figure}[t]
\begin{center}
\includegraphics[width=8.0cm]{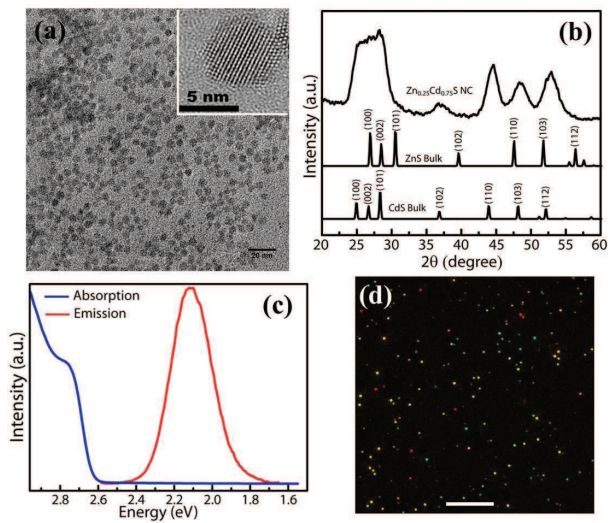}
\caption{(a) Transmission Electron Microscope (TEM) image of the Mn-doped Zn$_{0.25}$Cd$_{0.75}$S NCs of average size of 5.5 nm diameter, inset showing a high resolution TEM micrograph; (b) X-Ray Diffraction pattern, shown along with the standard bulk pattern of wurtzite ZnS and CdS; (c) Ensemble-averaged UV-visible absorption and photoluminescence spectra of the NCs dispersed in toluene show the strongly red-shifted Mn emission with respect to the band edge of the NC. The spectral features are what should be expected from a well-passivated Mn-doped semiconductor NCs, with a peak at ~ 587 nm (2.11 eV) with a width of about 70 nm (~ 250 meV), similar to a large number of reported Mn PL spectra; (d) Two color photoluminescence (PL) image of the NCs embedded in PMMA matrix, generated by a quantitative overlay of two separate images obtained with a green-yellow and an orange-red band-pass emission filters. The scale bar is 5 $\mu$m.}
\label{fig1}
\end{center}
\end{figure}

Among the most studied Mn-doped chalcogenide NCs, CdS is not suitable for the present investigation, since surface emissions from CdS host NCs often overlap the Mn emission ~\cite {24}, causing avoidable complications and the band gap of ZnS NCs, being in the UV-region, are not compatible with our excitation source.  In order to avoid these difficulties, we doped 0.9\% Mn in Zn$_{0.25}$Cd$_{0.75}$S alloyed NC hosts having wurtzite structure and an average diameter of 5.5 nm with large enough band gap of 465 nm (Fig. 1(a)-(c)) compatible with our experimental set-up. Single-particle PL imaging and spectroscopy from individual NCs have been performed on a very low ($\sim$ 1-2 nanomolar) concentration of these NCs embedded in polymethyl methacrylate (PMMA) thin film matrix by using a 457 nm Continuous Wave (CW) laser as excitation source through an epifluorescence microscope. All spectra were collected at room temperature, ensuring the usual level of participation from phonons in the emission process. All the individual diffraction-limited spots, as shown in Fig. 1(d), observed in our imaging show extensive PL intermittency (blinking), clearly demonstrating that we are close to the single particle detection limit in these measurements. Ab initio electronic structure calculations were performed including geometry optimization within a plane-wave pseudopotential approach using the VASP code ~\cite {30, 31}.

In order to probe the range of emission energies in the PL of these Mn-doped NCs, we have recorded individual energy-resolved emission spectra from more than a thousand image spots, and analyzed in detail 354 spots on the basis of their high signal-to-noise ratio. The crucial observation in our work is that single NC spectra are qualitatively different from what has been believed to be characteristic features of Mn emission from such doped nanocrystals. A representative set of normalized PL spectra is shown in Fig. 2, with peak positions varying from 2.34 eV (530 nm) to 1.97 eV (630 nm), representing emission colors ranging from green to red.
\begin{figure}[t]
\begin{center}
\includegraphics[width=8.0cm]{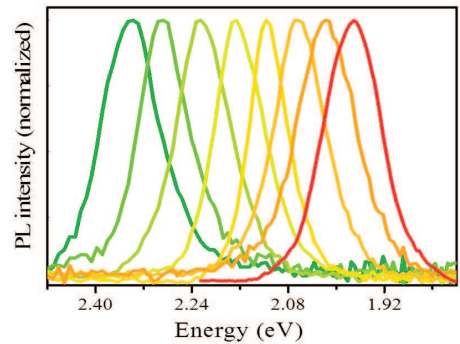}
\caption{Normalized Mn-emission spectra from eight individual NCs ranging from green to red color.}
\label{fig2}
\end{center}
\end{figure}
Not only the transition energies, but also the spectral width of PL emissions from individual NCs is strikingly different from that of the ensemble averaged one. This is illustrated in Fig. 3 where we compare the spectrum of a typical individual NC emitting close to 2.09 eV (593.6 nm) with the ensemble averaged spectrum of the same sample. It is evident that the emission line-width is significantly narrower (less than one-third) for the single NC as compared to that from a large collection of NCs.

We characterize the emission spectral features of each of the 354 single particle spectra in Fig. 4 by locating each spectrum in a two dimensional plot in the space of its peak position and the spectral width defined by the full width at half maximum (FWHM). These data make the following points obvious: Mn d-emission from Mn-doped semiconductor NCs can appear in wide range, contradicting the widely held belief that a PL emission involving entirely the dopant levels cannot be tuned over a significant energy window due to their atomic like, highly localized wave functions involved in the optical transition. It is fair to point out here that there have been sporadic attempts to tune ensemble-averaged Mn $\it {d}$-emission from similar systems achieving a tuning of the emission energy by 149 meV (~\cite {18}), 140 meV (~\cite {26}), 130 meV (~\cite {27}), and 95 meV (~\cite {28}), in contrast to a peak position spread of 370 meV shown in Figs. 2 and 4, thereby establishing a quantum jump in PL emission energy accessible via Mn-doping of such semiconductor NCs. Looking at the density distribution of data points in Fig. 4, we find that the most abundant occurrence of PL emission is within the window of 2.19 eV (565 nm) and 2.03 eV (610 nm) with a mean around 2.11 eV (587 nm), explaining why most of the ensemble averaged PL emission so far has been reported to be around 585 nm.
\begin{figure}[t]
\begin{center}
\includegraphics[width=8.0cm]{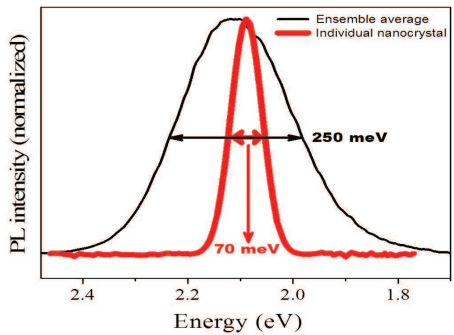}
\caption{PL spectrum of a single NC (thick line) compared with the ensemble averaged PL (thin line)}
\label{fig3}
\end{center}
\end{figure}

We find that the single-particle spectral line widths (FWHM) are distributed between a minimum of 60 meV (marked with an arrow in Fig. 4) and 147 meV. It is to be noted here that there cannot be any artefact of measurement that leads to a lowering of the FWHM below the intrinsic value. We recognize the possibility of the intrinsic FWHM of Mn $\it {d}$ PL emission being dependent on the emission energy. However, there is a large number of data points, spread over a wide window of the peak energy in Fig. 4, with FWHM $\leq$ 75 meV. One such spectrum can be seen in Fig. 3, exhibiting a symmetric line shape and 70 meV FWHM. Thus, it appears that the intrinsic spectral width of Mn emission from these NCs is in the order of 60-75 meV. Interestingly, PL spectra with larger spectral widths, in contrast, are often asymmetric, with a typical spectrum having an FWHM of 140 meV being shown as the inset A of Fig. 4. This spectrum with clearly perceptible shoulders on the red-edge establishes that this consists of multiple individual contributions. This spectrum is very well described in terms of three component spectra, each of which is symmetric with an FWHM of $\sim$ 75 meV. Similar analysis were carried out for most of individual data points in Fig. 4 that have FWHM larger than 75 meV and in each case, the PL spectrum is consistent with the idea of it being a combination of multiple contributions with narrow ($\leq$ 75 meV) width.  Thus, these results establish that the intrinsic FWHM of Mn $\it {d}$-emission in doped semiconductor NCs is $\leq$ 75 meV, and can be as low as 60 meV at the room temperature.
\begin{figure}[t]
\begin{center}
\includegraphics[width=8.0cm]{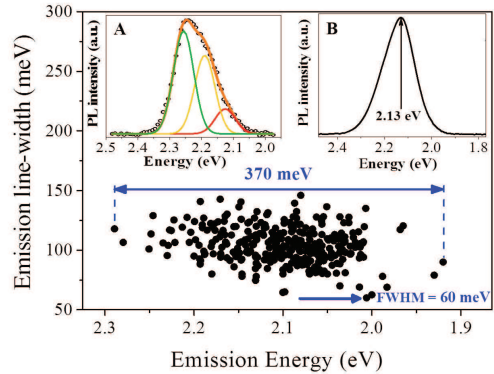}
\caption{PL line-width (FWHM) plotted against the corresponding emission energy for single NCs. The range of tunability ($\sim$ 370 meV) and the narrowest width ($\sim$ 60 meV) of PL emission are marked with arrows. Inset A shows a PL spectrum from a single NC, with an asymmetric shape due to multiple contributions. The experimental spectrum (open circles), and the fitted data (continuous orange line) and the three component spectra are shown in the figure. Inset B represents the sum of PL spectra from all individual NCs.}
\label{fig4}
\end{center}
\end{figure}
Interestingly, the sum of all individual single-NC PL spectra resembles that of the usually reported Mn emission PL spectrum with a well-defined peak at 2.13 eV (582 nm) (see inset B of Fig. 4). Thus, clearly ensemble averaged Mn PL spectrum is not representative of Mn emission from a NC, but is an artefact of averaging over a large number of contributions from different NCs, each having a distinct PL emission. There has been an attempt to attribute the earlier observed large width of the PL from Mn-doped systems ~\cite {29} to a coupling of the Mn electronic states to the vibronic properties of the host. Present results clearly negate this speculation; instead, it proves that the overlap of many PL contributions with different emission energies is the main reason for the large experimental width observed so far, as the experiments were performed in the ensemble averaging mode. Present results place an upper limit of $\sim$ 60-75 meV as the maximum contribution to the spectral width from such a vibronic coupling. The actual contribution may be even less, since Mn $\it {d}$-levels in the host are not exactly atomic-like due to a finite hybridization with the host electronic states, giving rise to a finite energy spread of Mn $\it {d}$ states.

Such a wide spread ($\sim$ 370 meV) of Mn emission energy from these samples is in apparent contradiction with the expected atomic-like ${\it {^4}}$${\it {T}}$${\it {_1}}$ - ${\it {^6}}$${\it {A}}$${\it {_1}}$ transition. To resolve this, we note that both multiplet states, ${\it {^6}}$${\it {A}}$${\it {_1}}$ and ${\it {^4}}$${\it {T}}$${\it {_1}}$, correspond to Mn $\it {d^{5}}$ electronic configurations. Within the ligand field theory, energies of these two multiplet configurations are given in terms of the Racah parameters A, B, and C or equivalently, the Slater F-integrals to express electron-electron interaction strengths and the crystal field parameter 10Dq, given by the energy difference between $\it {e}$ and ${\it {t_2}}$ states of Mn $\it {d}$. Ignoring the crystal field contribution for a moment, the energies of the ${\it {^6}}$${\it {A}}$${\it {_1}}$ state and the lowest energy ${\it {^4}}$${\it {T}}$${\it {_1}}$ state are (10A-35B) and (10A-25B+5C), respectively ~\cite {32}. Thus, Mn $\it {d}$ emission energy (= difference between the two multiplet energies) is given by (10B+5C). Interestingly, the emission energy is not a function of the parameter A, which is known to depend on the details of the specific system under investigation, as it is strongly influenced by screening that may change from system to system. In contrast, B and C are known to be insensitive to specific details of the system, being primarily governed by atomic properties of the Mn $\it {d}$ orbitals. Thus, any spread in the emission energy is not likely to arise from changes in the multi-electron Coulomb interactions within the Mn $\it {d}$-manifold; instead, it is most likely to arise from a change in the crystal field strength between different NCs. ~\cite {33} It is important to note here that there are a number of inequivalent sites for Mn doping in any semiconductor NC, due to the presence of the surface. In the present choice of the sample, the number of inequivalent sites is further increased due to the random substitution of Zn and Cd. This is likely to give rise to variations in the crystal field strength at the Mn site, depending on the specific substitution site realized in a given NC, thereby affecting the emission energy. In order to understand whether such variations in the crystal field strength at the Mn site can be strong enough to explain quantitatively the observed spread of 370 meV in the emission energy from different NCs (see Fig. 4), we have carried out first-principle electronic structure calculations for a given ZnS-CdS alloy NC of a fixed size with a single Mn substitution at various symmetry inequivalent cationic sites.
\begin{figure}[t]
\begin{center}
\includegraphics[width=8.0cm]{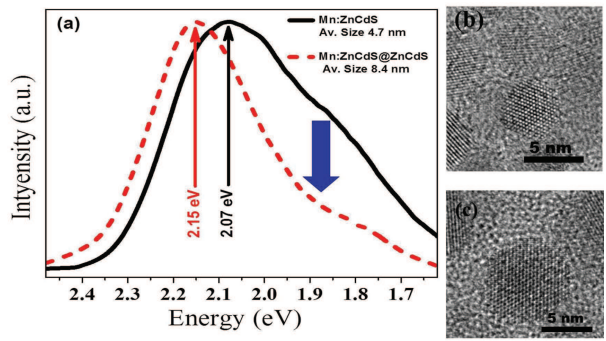}
\caption{Tuning of Mn-emission with size of Mn-doped ZnS-CdS alloyed nanocrystals illustrated with 4.7 nm and 8.4 nm sizes. The peak positions are indicated by thin vertical solid arrows. The thick arrow shows a preferential decrease of the PL intensity at the lower energy region on increasing the nanocrystal size, consistent with the idea that the lower energy emission arises from Mn species at and near the surface. (b) - (c) High Resolution Transmission Electron Microscopy images of the corresponding nanocrystals of average size 4.7 and 8.4 nm respectively.}
\label{fig5}
\end{center}
\end{figure}

It is to be noted that the pure crystal field splitting is considerably enhanced in a host by asymmetric hoppping between the host electronic states and Mn $\it {d}$-states, depending on whether it is a $\it {t_{2}}$ or an $\it {e}$ orbital of Mn. Thus, the relevant one-electron contribution to the transition energy between ${\it {^4}}$${\it {T}}$${\it {_1}}$ and ${\it {^6}}$${\it {A}}$${\it {_1}}$ states is best represented by the energy difference between the Mn $\it {t_{2}}$ and $\it {e}$ states, as it includes the effects of both the bare crystal field strength and shifts due to the hybridization with host states. This ligand field contribution to the transition energy in our various calculations for Mn at different sites was found to be spread over approximately 380 meV, in good agreement with the experimentally observed spread of 370 meV, with the ligand field splitting progressively increasing from the core towards the surface. Thus, it becomes clear that the extensive spread in the Mn PL emission energy observed by us, is dominated by changes in the local ligand field strength at the Mn site. Analyses of the calculated electronic structure show that this spread in the ($\it {t_{2}-e}$) energy difference is contributed by changes in the bare crystal field strength as well as by changes in the hopping strength between Mn $\it {d}$ and host $\it {s}$, $\it {p}$ states due to distortions in Mn-S bond-lengths and local symmetry.

These calculated results suggest that even the ensemble averaged Mn PL emission should be influenced, if one is able to change the ratio of the number of dopants nearer the core and that nearer the surface. Specifically, when Mn is doped in a smaller NC, the Mn $\it {d}$-emission is expected to have a higher contribution from the Mn-ions nearer to the surface as compared to those doped closer to the core. Since the ligand field splitting is smaller near the core, compared to that near the surface, thereby making the transition energy higher for the Mn doped near the core, one would therefore expect a reduction in the lower energy part of the ensemble averaged PL emission and a consequent blue-shift of the emission peak-position with the increasing size of the nanocrystal sample. We verified these expectations by making fresh samples of sizes 4.7 nm and 8.4 nm diameters (see Fig. 5), respectively smaller and larger than the 5.5 nm NC sample reported above. The PL peak positions of 2.07 eV (598 nm) and 2.15 eV (576 nm) for the 4.7 nm and the 8.4 nm samples, respectively (Fig. 5), as compared to 2.11 eV (587 nm) for the 5.5 nm sample are consistent with the interpretation of the spatially resolved data presented here in conjunction with the ab initio results. The fact that Mn$^{2+}$ ions experience different extent of ligand field strength at different sites may have important consequences in the magnetic and magneto-optical properties as well, since it influences both magnetocrystalline anisotropy and the dark-bright exciton splitting.

In conclusion, the present investigation on Mn-doped NCs establishes that Mn $\it {d}$-emission color may range from deep green to deep red, spanning an energy range of 370 meV. This contradicts the well-accepted belief in the community, with all studies so far reporting only a limited tuning of the transition energy ($\leq$ 150 meV) over the orange-red window. Our results suggest that the origin of this extraordinarily wide color tuning arises from a wide range of ligand field strengths at the Mn site, depending on the specific symmetry inequivalent, substitutional site occupied by the Mn ion in the NC host. In addition, the Mn PL emission is found to be very narrow ($\sim$ 60-75 meV FWHM) compared to all earlier publications reporting of an anomalously large line-width ($\geq$ $\sim$ 200 meV). Thus, in contrast to the published literature, this report shows that the Mn PL emission has an intrinsically narrow width with an unprecedented possibility of tuning the emission color from the deep red to deep green.\\

\noindent \textbf{Synthesis of Mn-doped Zn$_{0.25}$Cd$_{0.75}$S nanocrystals}

\noindent For a typical synthesis of Mn-doped Zn$_{0.25}$Cd$_{0.75}$S NCs, the reaction mixture consisting of 0.289 mmol of CdO (s.d. fine-chemicals Limited), 0.097mmol of ZnO (Leo Chemical), 1 mL of oleic acid (Aldrich), and 10 mL of 1-octadecene (Aldrich) was degassed with nitrogen at 150 $^{o}$C for 30 min. It was then heated up to 310 $^{o}$C giving a clear solution. 0.0038 mmol of Mn(CH3COO)$_{2}$.4H$_{2}$O in oleyl amine (Aldrich, 1 mL) was injected to the above hot solution. Consequently, a solution of S (1.9 mmol in 1 mL of 1-octadecene) was injected followed by growth at 310 $^{o}$C for 20 min. All the steps have been carried out in a nitrogen atmosphere, and the product NCs were precipitated and washed repeatedly with 1-butanol. The washed NCs after drying under vacuum, dispersed in toluene for optical measurements.\\

Authors thank Department of Science and Technology and Board of Research in Nuclear Sciences, Government of India, for financial support. AH thanks CSIR for financial support. AC, AL and SD thank CSIR and IRCC, IIT Bombay for PhD fellowship and financial support.\\

${*}$ Also at Jawaharlal Nehru Centre for Advanced Scientific Research, Bangalore, India.
sarma@sscu.iisc,ernet.in


\begin{thebibliography}{}

\bibitem {1} D. Langer, and S. Ibuki, Phys. Rev. {\bf 138}, A809 (1965).

\bibitem {2} D. W. Langer, and H. J. Richter, Phys. Rev. {\bf 146}, 554 (1966).

\bibitem {3} R. Parrot, and C. Blanchard, Phys. Rev. B {\bf 6}, 3992 (1972).

\bibitem {4} H. -E. Gumlich, J. Lumin. {\bf 23}, 73 (1981).

\bibitem {5} D. D. Thong, and O. Goede, Phys. Stat. Sol. b {\bf 120}, K145 (1983).

\bibitem {6} J. H. Yu, X. Liu, K. E. Kweon, J. Joo, J. Park, K. T. Ko, D. W. Lee, S. Shen, K. Tivakornsasithorn, J. S. Son, J. H. Park, Y. W. Kim, G. S. Hwang, M. Dobrowolska, J. K. Furdyna, and T. Hyeon, Nature Mat. {\bf 9}, 47 (2010).

\bibitem{7} P. M. Koenraad, and M. E. Flatte, Nature Mat. {\bf 10}, 91 (2011).

\bibitem{8} R. Viswanatha, J. M. Pietryga, V. I. Klimov, and S. A. Crooker, Phys. Rev. Lett. {\bf 107}, 067402 (2011)

\bibitem{9} C. L. Gall, A. Brunetti, H. Boukari, and L. Besombes, Phys. Rev. Lett. {\bf 107}, 057401 (2011).

\bibitem{10} L. Besombes, Y. Leger, L. Maingault, D. Ferrand, and H. Marriette, Phys. Rev. Lett. {\bf 93}, 207403 (2004).

\bibitem{11} Y. Leger, L. Besombes, L. Maingault, D. Ferrand, and H. Marriette, Phys. Rev. Lett. {\bf 95}, 047403 (2005).

\bibitem{12} M. Goryca, T. Kazimierczuk, M. Nawrocki, A. Golnik, J. A. Gaj, and P. Kossacki, Phys. Rev. Lett. {\bf 103}, 087401 (2009).

\bibitem{13} R. N. Bhargava, D. Gallagher, X. Hong, and  A. Nurmikko,  Phys. Rev. Lett. {\bf 72}, 416 (1994).

\bibitem{14} A. A. Bol, and A. Meijerink, Phys. Rev. B {\bf 58}, R15997 (1998).

\bibitem{15} S. C. Erwin, L. J. Zu, M. I. Haftel, A. L. Efros, T. A. Kennedy, and D. J. Norris, Nature {\bf 436}, 91 (2005).

\bibitem{16} G. M. Dalpian, and J. R. Chelikowsky, Phys. ReV. Lett. {\bf 96}, 226802 (2006).

\bibitem{17} P. I. Archer, S. A. Santangelo, and D. R. Gamelin, Nano Lett. {\bf 7}, 1037 (2007).

\bibitem{18} S. Ithurria, P. Guyot-Sionnest, B. Mahler, and B. Dubertret, Phys. Rev. Lett. {\bf 99}, 265501 (2007).

\bibitem{19} C. Gan, Y. Zhang, D. Battaglia, X. Peng, and M. Xiao, Appl. Phys. Lett. {\bf 92}, 241111 (2008).

\bibitem{20} M. Goryca, P. Plochocka, T. Kazimierczuk, P. Wojnar, G. Karczewski, J. A. Gaj, M. Potemski, and P. Kossacki, Phys. Rev. B {\bf 82}, 165323 (2010).

\bibitem{21} A. Nag, S. Chakraborty, and D. D. Sarma, J. Am. Chem. Soc. {\bf 130}, 10605 (2008).

\bibitem{22} T. L. Chan, A. T. Zayak, G. M. Dalpian, and J. R. Chelikowsky, Phys. Rev. Lett. {\bf 102}, 025901 (2009).

\bibitem{23} A. A. Gunawan, K. A. Mkhoyan, A. W. Wills, M. G. Thomas, and D. J. Norris, Nano Lett. {\bf 11}, 5553 (2011).

\bibitem{24} A. Nag, and D. D. Sarma, J. Phys. Chem. C {\bf 111}, 13641 (2007).

\bibitem{25} D. V. Talapin, I. Mekis, S. Gotzinger, A. Kornowski, O. Benson, and H. Weller, J. Phys. Chem. B {\bf 108}, 18826 (2004).

\bibitem{26} N. Pradhan, and X. G. Peng, J. Am. Chem. Soc. {\bf 129}, 3339 (2007).

\bibitem{27} A. Nag, R. Cherian, P. Mahadevan, A. V. Gopal, A. Hazarika, A. Mohan, A. S. Vengurlekar, and D. D. Sarma, J. Phys. Chem. C {\bf 114}, 18323 (2010).

\bibitem{28} R. S. Zeng, T. T. Zhang, G. Z. Dai, and B. S. Zou, J. Phys. Chem. C {\bf 115}, 3005 (2011).

\bibitem{29} R. Beaulac, P. I. Archer, S. T. Ochsenbein, and D. R. Gamelin, Adv. Funct. Mater. {\bf 18}, 3873 (2008).

\bibitem{30} G. Kresse, and J. Furthmuller, Phys. Rev. B {\bf 54}, 11169 (1996).

\bibitem{31} G. Kresse, and D. Joubert, Phys. Rev. B {\bf 59}, 1758 (1999).

\bibitem{32} J. S. Griffiths, The Theory of Transition Metal Ions. Cambridge University Press: Cambridge, U.K., {\bf 1971}.

\bibitem{33} Here we have considered only single Mn ion doped in a single nanocrystal. From our preliminary results for multiply doped nanocrystals, we find that Mn ions remain far apart from each other for energetic reasons. The change in Mn transition energy due to Mn-Mn interaction is appreciable only for two Mn ions in the nearest neighbour cationic sites. The Mn-Mn interaction is found to be suppressed to the point that the transition energy is affected by less than 10 meV even with two Mn ions separated by only 6.7 $\sim$ 13 {\AA}, while we estimate the average Mn-Mn separation to be $\sim$ 13 {\AA} in our experiments.



\end{thebibliography}
\end{document}